\documentclass[aps,pre,twocolumn,superscriptaddress]{revtex4}

\usepackage{graphicx}
\usepackage{amssymb}
\usepackage{amsmath}
\usepackage[british]{babel}

\begin{document}

\title{Particle invasion, survival, and non-ergodicity in 2D diffusion processes
with space-dependent diffusivity}

\author{A. G. Cherstvy}
\affiliation{Institute for Physics \& Astronomy, University of Potsdam,
14476 Potsdam-Golm, Germany}
\author{A. V. Chechkin}
\affiliation{Institute for Physics \& Astronomy, University of Potsdam,
14476 Potsdam-Golm, Germany}
\affiliation{Institute for Theoretical Physics, Kharkov Institute of Physics
and Technology, Kharkov 61108, Ukraine}
\affiliation{Max-Planck Institute for the Physics of Complex Systems,
N{\"o}thnitzer Stra{\ss}e 38, 01187 Dresden, Germany}
\author{R. Metzler}
\affiliation{Institute for Physics \& Astronomy, University of Potsdam,
14476 Potsdam-Golm, Germany}
\affiliation{Department of Physics, Tampere University of Technology, 33101
Tampere, Finland}

\date{\today}

\begin{abstract}
We study the thermal Markovian diffusion of tracer particles in a 2D medium with
spatially-varying diffusivity $D(r)$, mimicking recently measured, heterogeneous
maps of the apparent diffusion coefficient in biological cells. For this
heterogeneous diffusion process (HDP) we analyse the mean squared displacement
(MSD) of the tracer particles, the time averaged MSD, the spatial probability
density function, and the first passage time dynamics from the cell boundary to the
nucleus. Moreover we examine the non-ergodic properties of this process which are
important for the correct physical interpretation of time averages of observables
obtained from single particle tracking experiments. From extensive computer
simulations of the 2D stochastic Langevin equation we present an in-depth study of
this HDP. In particular, we find that the MSDs along the radial and azimuthal
directions in a circular domain obey anomalous and Brownian scaling, respectively.
We demonstrate that the time averaged MSD stays linear as a function of the lag
time and the system thus reveals a weak ergodicity breaking. Our results
will enable one to rationalise the diffusive motion of larger tracer particles
such as viruses or submicron beads in biological cells.
\end{abstract}

\maketitle

\section{Introduction}

For a typical bacterial cell such as \textit{E. coli\/}, various proteins, large
cellular complexes, nucleic acids, lipids, etc. occupy some 30-40\% of the cell
volume \cite{crowd1,crowd2,crowd3,crowd4}. The implications of this macromolecular
crowding on the characteristics of diffusing particles of various sizes are still
under debate \cite{franosch13,pt}. Another source impeding the free diffusion of
particles in eukaryotic cells stems from a network of cytoskeletal filaments and
internal membranes like the endoplasmic reticulum or the nuclear membrane. Such
forms of crowding impair the particle diffusivity inside a cell and may alter the
law of diffusion altogether, from Brownian motion to a subdiffusive law. In the
latter case, the mean squared displacement (MSD) scales as \cite{report}
\begin{equation}
\left<x^2(t)\right>\simeq t^\beta,
\end{equation}
with the anomalous diffusion exponent $0<\beta<1$. Experimental data are available,
inter alia, for the in vivo subdiffusion of proteins \cite{lang-nucleus} and enzymes
\cite{engelborghs}, endogenous submicron particles (lipid and insulin granules)
\cite{elbaum,lene04,jeon11,tabei,naturephot}, viral particles \cite{brauch01},
fluorescently labelled gold particles \cite{guigas},
messenger RNA molecules \cite{goldingcox}, as well as the telomeres of chromosomes
\cite{telosubdiff}. In vitro, dense solutions of coil-like polymers, proteins, or
worm-like micelles often mimic the effects of molecular crowding which depend on
the particle size, the solution viscosity, and the effective medium porosity
\cite{pan,weiss,lene,fradin05}. Similarly, in large scale computer simulations of
crowded lipid membranes, subdiffusion is observed for various membrane chemistries
\cite{kneller,jeon,akimoto}.

\begin{figure}
\includegraphics[width=8cm]{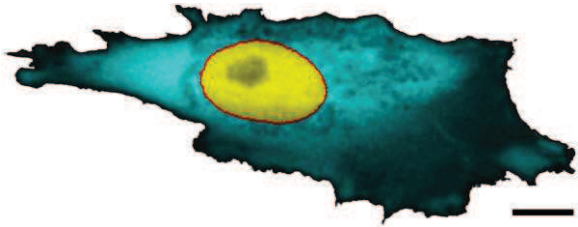}
\caption{The variation of local diffusivity in the cytoplasm of a mammalian
cell. The FRAP intensity of the cyan colour refers to the local effective
porosity of the cytoplasm, scaling with the volume fraction available
for protein diffusion. Scale bar is 10 $\mu$m. The image is taken from
Ref. \cite{lang11}; courtesy to J\"org Langowski.}
\label{fig-diff-lang}
\end{figure}

Measuring the apparent \emph{local\/} diffusivity of smaller proteins in bacterial
\cite{elf12} and eukaryotic \cite{lang11} cells reveals a nontrivial dependence on
the position in the cell. One reason for this spatial variation of the diffusivity
may be the cells' geometrical shape \cite{elf12}. Thus, certain cell types possess
a `fried egg-shape' (Fig.~\ref{fig-diff-lang}) with a significant variation of the
cell thickness from the periphery towards the nucleus. A higher apparent abundance
of proteins in the cytosol near the nucleus, interpreted as a higher cytoplasm
diffusivity, may simply originate due to the 2D imaging of the fully 3D particle
trajectories. Away from the thicker perinuclear region, the cell periphery offers
only a thin, nearly 2D domain for the particle diffusion.

Another source for the
variations of the local diffusivity is the heterogeneity of the density of the
macromolecular crowding in the cytoplasm and nucleoplasm, as well as of the dense
cytoskeletal meshwork near the cell periphery, and the accumulation of large
cellular organelles in a perinuclear region. How exactly this affects the porosity
of the cytoplasm and the diffusivity of tracers of different sizes is not well
established \cite{kuhn-comments,poland11}.
Specifically, substantial deviations from the Stokes-Einstein law for protein
tracers of varying molecular weights (MW) diffusing in the \textit{E. Coli\/}
cytoplasm were observed and the diffusivity shown to follow the scaling law $D
\sim\mathrm{MW}^{-0.7}$ \cite{diff-mw}. Small tracer proteins apparently experience
a higher porosity near the nucleus of mammalian cells \cite{lang11}, while the
diffusion of larger proteins becomes progressively restricted \cite{diff-mw}.
Thus, from a biological perspective,
a stochastic model with spatially-varying diffusivity may mimic the effects on
the diffusion of tracer particles in the heterogeneous environment of the crowded
cellular cytoplasm and will serve as an empirical description of secondary
processes such as intracellular, diffusion-controlled reactions. We here study
the physical properties of such a heterogenous diffusion process (HDP).

Important clues come from viral particles, a major class of natural diffusers
in the bacterial cytoplasm. After internalisation by receptor-driven endocytosis
\cite{endo1}, viruses often recruit highly processive cellular motor proteins
\cite{sodeik1,sodeik2} which ensure fast and efficient viral transport between the
cell periphery and the nucleus, where the viral replication and assembly often
occurs \cite{trafik1,trafik2}. The intra-cellular dynamics of viruses and their
multi-step infection pathways, as monitored by single-particle tracking, exhibits
some features of anomalous diffusion \cite{brauch01}. For instance, the scaling
exponent $\beta$ of the viral motion is shown to depend on the region of the
cytoplasm in which the diffusion takes place.

Three different modes of transport for adeno-associated viruses \cite{samulMT,
AAV-traffik} were identified in living substrate-adhered HeLa cells \cite{brauch01,
seisenphd,brauch02}. The first is Brownian motion with $\beta$=1 albeit with a
much smaller diffusion coefficient than in dilute aqueous solution. The second
mode is that of subdiffusion with $\beta=0.5\ldots0.9$ and with a broad apparent
distribution of diffusivities. The third mode is that of motor-driven transport
of viruses via quasi-1D persistent walks along microtubular filaments, mediated
by molecular motors driven by energy from ATP conversion. Upon infection, the
ballistic, driven motion with $\beta$=2 yields an effective drift of virions
towards the nucleus. These diffusion-based and active modes of viral transport can
interchange. Although the fraction of actively-transported virions is
relatively small \cite{brauch01}, this `active pathway' is often vital for
a successful viral infection. Small viruses can reach the nucleus solely by
thermal diffusion, while larger virions have no chance but rely on the active
transport mechanism. Indeed, an accumulation of viruses near the nucleus was
shown to be inhibited by microtubuli-depolymerising drugs (e.g., nocodazole)
that suppress motor-assisted virus transport \cite{mitra05}.

Several recent models of intermittent transport \cite{bress13,gryb} were
implemented to describe  kinetics of viral infection, and search optimisation
models with 2D versus 3D intermittent dynamics were developed \cite{intermitt-first,
intermitt-last}. A number of diffusion \cite{bress11}, diffusion-reaction-advection
\cite{mitra05}, and kinetic transport \cite{smith01} models were suggested to
rationalise the features of intracellular virus trafficking. In particular, the
kinetics of spreading of a viral population starting at the cell membrane and the
accompanying nucleus invasion times were computed \cite{smith01} and compared to
typical time scales of viral infection recorded experimentally
\cite{sodeik-temporal,japan-temporal}.
In a series of theoretical and computer simulation studies Holcman and colleagues
\cite{holc07,holc08,holc09,holc092,holc12} modelled the process of
viral trafficking as a sequence of alternating Brownian 2D diffusive excursions
and ballistic motor-powered propulsions along radially-ordered microtubuli
filaments. The dynamical characteristics of viral invasion were computed in
such a 2D planar pie-like model. The probability density function (PDF) and
the mean time of nucleus invasion by viruses were evaluated \cite{holc09}. More
advanced theoretical models can also include a rate of viral degradation in
the cytoplasm \cite{smith01}, the kinetics of viral binding to microtubuli,
and some bi-directionality of virus transport by the motors.

Here we consider the passive diffusion of tracer particles of sizes comparable
to a virus capsid in a model cell. To construct our model we include the following
information.
From the viral trajectories reported in Refs.~\cite{brauch01,seisenphd} we
conclude that those particles exhibiting normal diffusion with $\beta=1$ take
azimuthal journeys, at about constant separation from the cell nucleus. Such
propagation likely takes place in a region of roughly constant diffusivity. In
contrast, that part of the viral population that diffuses anomalously mainly
travels in the radial direction. We propose below that heterogeneities of
the medium during the journey of a particle from the cell membrane to the
nucleus gives rise to anomalous (in particular, sub-diffusive) features
for this second population of particles.

Recently, extending previous studies \cite{polska-Dx-MSD,fedotov-hetero} we
examined effects of position-dependent diffusivities on the ensemble and time
averaged characteristics for 1D HDPs \cite{hdp13a,hdp13b}. We
tested several functional forms for $D(x)$ (power-law, logarithmic, and
exponential) to rationalise their implications onto diffusive and ergodic
properties of the process. For power-law forms of $D(x) $ we predicted from
stochastic simulations and analytical calculations the regimes of sub- and
super-diffusive behaviour. The conditions for weak ergodicity breaking were
also analysed in details, an important feature when information from single
particle tracking studies is evaluated in terms of time averages \cite{pt}.
The diffusion is non-ergodic in 1D due to the heterogeneities of the medium.
In general, despite a non-Brownian scaling of the MSD, the time averaged MSD
was shown to follow a strictly linear growth with lag time. These
features are similar to those for continuous time random walk processes
\cite{rm-ctrw}.

For the 2D HPDs examined below we also find non-ergodic behaviour. In addition,
we compute a number of biologically relevant quantities such as the survival
probability $S(t)$ of particles in a circular domain for both diffusion from
the inside of the cell to the outside and vice versa, the first-passage time
dynamics for reaching the domain boundary, the PDF for the spreading of diffusing
particles starting at the cell centre, at the cell boundary, and for initially
uniformly-distributed walkers.

This paper is organised as follows. In Section \ref{sec-model} we introduce
the basic notations and the main quantities to be analysed. We outline
the numerical scheme used in computations as well as the implemented theoretical
concepts. In Section \ref{sec-results} we report the main simulations results
and support them by asymptotic analytic calculations. We analyse the effects of the
system heterogeneity and polar cell geometry on diffusive, kinetic, and ergodic
properties of the HDP. In Section \ref{sec-discussion} the Conclusions are
drawn and possible applications of the results are discussed.

\section{Model}
\label{sec-model}
   
Our model cell is a circular disc with a reflecting outer boundary, mimicking
the situation that internalised viruses do not leave the cell again. We analyse
the following form of the diffusivity 
\begin{equation}
D(r)=D_0\frac{A}{A+r^2},
\label{eq-Dr}
\end {equation}
that is solely dependent on the radius $r$ away from the cell centre. At small
$r$ values the diffusion is fastest, continuously slowing down towards the
outer cell region (the `cell membrane'). To avoid divergencies in the discrete
simulations scheme implemented below, we regularised $D(r)$ in Eq.~(\ref{eq-Dr})
by introduction of the constant $A>0$. At $r\gg A$, the diffusivity exhibits the
power-law scaling $D(r)\sim1/r^2$. The constant $D_0$ fixes the units of the
diffusivity. The dependence (\ref{eq-Dr}) is in qualitative agreement with the
experimentally measured trends for the diffusivity of small fluorescently-labelled
proteins in the cytoplasm of mammalian NLFK and HeLa cells \cite{lang11}. It also
reflects the above observation that azimuthal diffusion is fully Brownian, i.e.,
in our language, the diffusivity remains constant. The simulations method described
below is readily applicable to other $D(r)$ forms.

We characterise the HDP in terms of the ensemble-averaged MSD of particles defined
via the PDF $P(r,t)$,
\begin{equation}
\label{eq-r2-msd}
\left<r^2(t)\right>=\int r^2P(r,t)2\pi rdr.
\end {equation}
For a 2D trajectory $\mathbf{r}(t)=\{x(t),y(t)\}$ ($r=\sqrt{\mathbf{r}^2}$) of
length $T$, the time averaged MSD is defined as the sliding average with the lag
time $\Delta$,
\begin{eqnarray}
\label{TAMSD}
\nonumber
\overline{\delta^2(\Delta)}&=&\frac{1}{T-\Delta}\int\limits_0^{T-\Delta}
\Big(\left[x(t+\Delta)-x(t)\right]^2\\
&&\hspace*{1.6cm}+\left[y(t+\Delta)-y(t)\right]^2\Big)dt.
\end{eqnarray}
While for an ergodic process for sufficiently long measurement times $T$ the
equivalence $\left<r^2(\Delta)\right>=\overline{\delta^2(\Delta)}$ holds, the
behaviour of the two quantities remains different even for $T\to\infty$ in
weakly non-ergodic systems \cite{pt,rm-ctrw,pccp}. In particular, individual
realisations of time averaged quantities becomes irreproducible \cite{pt,rm-ctrw,
pccp}.

The ergodicity breaking parameter $\mathrm{EB}$ characterises the deviation
of the system from the ergodic behaviour. It contains the second moment of the time
averaged MSD and is defined as follows \cite{barkai-eb,russians-eb}
\begin{equation}
\label{EB-2D}
\mathrm{EB}(\Delta)=\lim_{T/\Delta\to\infty}\frac{\left<\left(\overline{\delta
(\Delta)^2}\right)^2\right>-\left<\overline{\delta(\Delta)^2}\right>^2}{\left<
\overline{\delta(\Delta)^2}\right>^2}.
\end{equation}
For the canonical Brownian motion in 1D ($d=1$) one obtains \cite{barkai-eb}
\begin{equation}
\label{eb-bm}
\mathrm{EB_{BM}}(d=1,\Delta)=\frac{4}{3}\frac{\Delta}{T}.
\end {equation}
This means that $\mathrm{EB_{BM}}\to0$ at $\Delta/T\to0$, and the spread of time
averaged MSD traces around the mean computed over $N$ traces,
\begin{equation}
\label{TAMSD-mean}
\left<\overline{\delta^2(\Delta)}\right>=N^{-1}\sum_{i=1}^N\overline{\delta^2_i
(\Delta)},
\end{equation}
approaches a sharp $\delta$-function shape, i.e., the experiment is fully
reproducible \cite{pt,rm-ctrw,pccp}. To extract a
statistically meaningful spread of $\overline{\delta^2 (\Delta)}$ values around
the mean $\left<\overline{\delta^2}\right>$, the condition $\Delta/T\ll1$ should
be satisfied.

We also define the survival probability $S(t)$ of particles in the circular domain
when either of the boundaries is considered absorbing, and the particles are
released at the opposite boundary. The probability of particles in this scenario
is not conserved and $S(t)$ tends to zero as time progresses. The PDF of first
passage is then defined as $-dS(t)/dt$, and the mean first passage time as
$\mathrm{MFPT}=\int_0^{\infty}S(t)dt$. We evaluate the statistics of the first
arrival times directly from the generated trajectories, $\mathbf{r}(t)$.

At every time step in the computer simulations we use the Klimontovich-H{\"a}nggi
\cite{haenggi}
post-point scheme to evaluate the two coupled Langevin equations with independent
noise sources,
\begin{eqnarray}
\nonumber
x_{i+1}-x_i=\sqrt{2D\left(\sqrt{x^2_{i+1}+y^2_{i+1}}\right)}(W_{x,i+1}-W_{x,i}),\\
y_{i+1}-y_i=\sqrt{2D\left(\sqrt{x^2_{i+1}+y^2_{i+1}}\right)}(W_{y,i+1}-W_{y,i}).
\label{langevin-simul-2D}
\end {eqnarray}
Here, the increments of the Wiener processes for the corresponding coordinate,
$(W_{x,i+1}-W_{x,i})$ and $(W_{y,i+1}-W_{y,i})$, each represent a different
$\delta$-correlated Gaussian noise with unit variance. Unit time intervals
$\delta t$ separate consecutive iteration steps in the simulations. From $N$
2D stochastic trajectories $\{x(t),y(t)\}$ generated for the initial particle
position, $x(t=0)=x_0$ and $y(t=0)=y_0$, the ensemble and time averaged
characteristics of the HDP are evaluated. We note that we could also use the
Stratonovich scheme to simulate the process. For the MSD, similar to the 1D case
\cite{hdp13a}, the difference between the two representations occurs only in the
prefactor and is of order unity.

\section{Results}
\label{sec-results}

\subsection{MSD and time averaged MSD}

\begin{figure}
\includegraphics[width=8cm]{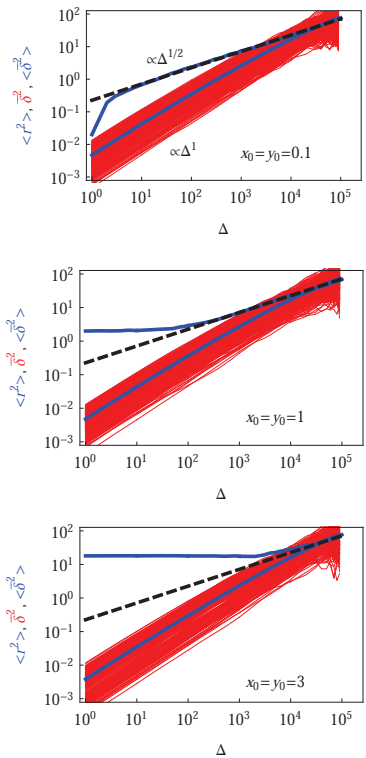}
\caption{Dependence of the ensemble-averaged MSD (thick blue curve), the mean
time averaged MSD (thick blue curve), and the time averages of the MSD for
individual trajectories (red curves) on time $t$ or the lag time $\Delta$. The
theoretical asymptote (\ref{eq-strat}) for the ensemble-averaged MSD is represented
by the dashed black line. Parameters: $A=0.01$, the starting positions are $x_0=
y_0=0.1$, 1, and 3 from the graphs from top to bottom. The number of traces for the
averaging is $N=300$, the length of each trajectories being $T=10^5$ in units of
the simulations time step. The simulation time for each choice of the starting
conditions is $\sim$2.5 days on a standard 3 GHz working station.}
\label{fig-msd-tamsd-out}
\end{figure}

The computed ensemble and time averaged MSD as well as the mean time averaged MSD
are shown in Fig.~\ref{fig-msd-tamsd-out}. For the 1D case, the MSD for
a diffusivity of the form $D(x)=D_0|x|^2$ reveals the subdiffusive scaling
\begin{equation}
\left<x^2(t)\right>\approx4\pi^{-1/2}\sqrt{D_0At}\simeq t^{1/2}.
\label{eq-strat}
\end {equation}
This asymptote, derived within the Stratonovich scheme in Ref.~\cite{hdp13a} and
shown as the black dashed curves in Fig.~\ref{fig-msd-tamsd-out}, is in good
agreement with our 2D simulations for the $D(r)$ defined in Eq.~(\ref{eq-Dr}).
At proximate initial positions $x_0$ and $y_0$, the deviations from the theoretical
MSD asymptote (\ref{eq-strat}) almost vanish after several simulation steps. For
more distant initial positions $\{x_0,y_0\}$, the sub-linear $\left<r^2(t)\right>
\simeq t^{1/2}$-scaling is approached somewhat later, giving rise to an initial
plateau.

The  time averaged MSD trajectories are \emph{linear\/} functions of the lag
time $\Delta$, their mean scaling as
\begin{equation}
\left<\overline{\delta^2(\Delta)}\right>\simeq\Delta^1,
\end{equation}
see Fig.~\ref{fig-msd-tamsd-out}. The spread (amplitude scatter) of the time
averaged MSD traces is very pronounced, with large trajectory-to-trajectory
variations, see the red traces in Fig.~\ref{fig-msd-tamsd-out}. This indicates
an ergodic violation, see below. Also, at shorter $T$ values the spread of
individual $\overline{\delta^2(\Delta)}$ in the region of $\Delta/T\ll1$
progressively decreases for larger values of the particle initial position (not
shown). Here, we do not quantify the details of the distribution $\phi(\overline{
\delta^2}/\langle\overline{\delta^2}\rangle)$ of individual traces $\overline{
\delta^2}$. We refer the reader to Refs.~\cite{hdp13a,hdp13b} where this procedure
is discussed in detail for the 1D HDP.

We checked that for $D(x)=const$ we obtain the standard 2D result with
\begin{equation}
\label{eq-MSD-4Dt}
\left<r^2_{\text{BM}}(t)\right>=4D_0t,
\end{equation}
with only a minute scatter of $\overline{\delta^2(\Delta)}$ traces at $\Delta/T
\ll1$. Moreover, we find that
\begin{equation}
\mathrm{EB}_{\text{BM}}(d=2,\Delta)=\frac{\mathrm{EB_{BM}}(d=1,\Delta)}{2}, 
\label{eq-eb-d2}
\end{equation}
indicative of the self-averaging behaviour typical for Brownian motion. For the
choice of $D(r)$ used here, leading to subdiffusion, usually the ratio $\left<
\overline{\delta^2(\Delta)}\right>/\left<r^2(\Delta)\right>\ll1$ for not too small
values of the initial positions $\{x_0,y_0\}$, see Fig.~\ref{fig-msd-tamsd-out}.

As a connection to experiments, let us define the model parameters that can
describe the MSD magnitudes measured in the tracking experiments of small
adeno-associated viruses \cite{brauch01,seisenphd,brauch02}, as mentioned in the
Introduction. Specifically, for
the subdiffusive population of viruses the MSD measured in the cells after the
diffusion time of $t\approx0.32$s was $\approx0.4\mu\mathrm{m}^2$, see Fig.~3G in
Ref.~\cite{brauch01}. The viral diffusivity was $D\sim0.2\mu\mathrm{m}^2
\mathrm{s}^{-0.6}$ for their subdiffusive motion with exponent $\beta\approx0.6$.
To get the same MSD value in the same physical time $t$ the diffusion
coefficient $D_0$ in our model would be $D_0\approx10\mu\mathrm{m}^4\mathrm{s}^{
-1}$.

\subsection{Azimuthal and radial diffusion}
        
\begin{figure}
\includegraphics[width=7cm]{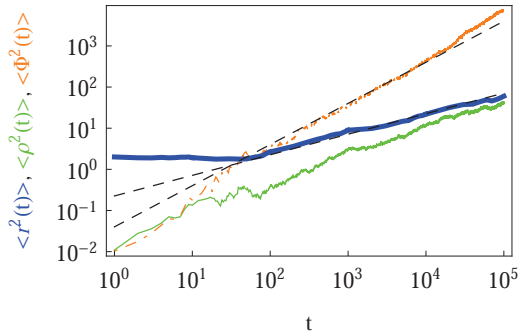}
\caption{Anomalous behaviour of radial increments (orange) and Brownian diffusion
of azimuthal particle increments (green line). The dashed asymptotes are given by
Eqs.~(\ref{eq-strat}) and (\ref{eq-MSD-4Dt}). The MSD corresponds to the blue line.
Parameters are the same as in Fig.~\ref{fig-msd-tamsd-out}, except for $x_0=y_0=1$
and $N=40$.} 
\label{fig-radius-phase}
\end{figure}

We project the increments of the diffusing particles at each simulation step $i$
with particle position $r_i$ onto the radial and azimuthal directions and compute
the single-step displacements $\delta r_i$ and $r_i\delta\Phi_i$. We account for
the clock- and anti-clockwise azimuthal rotation of the particle position vector
$\textbf{r}_i$. We then restore the corresponding average displacements after $t
=T/\delta t$ simulation steps, computed as the average over all the traces, $\left<
\rho^2(t)\right>=\left<(\sum_{i=1}^{t}\delta r_i)^2\right>$ and $\left<\Phi^2(t)
\right>=\left<(\sum_{i=1}^{t}r_i\delta\Phi_i)^2\right>$. The results of the
simulations show that the growth of the radial increments, similarly to the MSD
in Eq.~(\ref{eq-strat}), obeys the subdiffusive law
\begin{equation}
\left<\rho^2(t)\right>\simeq t^{1/2}.
\end{equation}
For the azimuthal increments, in contrast, the diffusion is Brownian,
Fig.~\ref{fig-radius-phase}, with the scaling
\begin{equation}
\left<\Phi^2(t)\right>\approx4D_0t.
\end{equation}

\subsection{Ergodic violation}
     
The simulations show that the ergodicity breaking parameter for short lag times
$\Delta$ assumes values close to those for the 1D case with analogous $D(x)$
treated in Ref.~\cite{hdp13a}. The EB values at $\Delta/T\ll1$ deviate from zero,
indicating a weak ergodicity breaking and non-equivalence of ensemble and time
averaging for this 2D diffusion process in a heterogeneous environment.
Non-homogeneities in the diffusion coefficient break the ergodicity in the system,
see also the discussion in Ref.~\cite{inhomo-nonergodic}.

For long lag times, when $\Delta\to T$, and for $\{x_0,y_0\}$ values in the
high-diffusivity region close to $r=0$, the ergodicity breaking parameter
approaches 1/2 of the value (\ref{eb-bm}) of the asymptote for 1D Brownian motion
obtained in Ref.~\cite {barkai-eb}. Such a reciprocal dependence on the space
dimension $d$,
\begin{equation}
\label{eq-EB-d}
\text{EB}(d,\Delta)=\frac{\text{EB}(1,\Delta)}{d},
\end{equation}
has recently also been discovered for multi-dimensional fractional Brownian motion
\cite{rm-eb-d}.

\begin{figure}
\includegraphics[width=7cm]{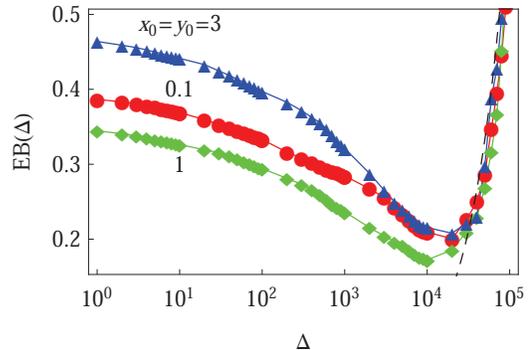}
\caption{Ergodicity breaking parameter as a function of lag time $\Delta$ for
different initial conditions. The Brownian asymptote (\ref{eq-eb-d2})
for the 2D case is shown by the dashed line. Parameter are the same as in
Fig.~\ref{fig-msd-tamsd-out}.}
\label{fig-eb}
\end{figure}

Clearly, in our system with an inhomogeneous diffusivity, the initial conditions
of the diffusing particles affect the magnitude of the time averaged MSD traces
and thus the values of the ergodicity breaking parameter. Specifically, as the
values of $x_0$ and $y_0$ decrease, the value of $\text{EB}(\Delta\to1)$ decreases,
as shown in
Fig.~\ref{fig-eb}. The HDP thus becomes more ergodic, and the Brownian asymptote
(\ref{eq-eb-d2}) is approached at earlier lag times $\Delta$. This is due to the
fact that at larger $\{x_0,y_0\}$ the spatial heterogeneities are sampled by the
considerably slower walkers to a lesser extent for the same length $T$.

Note here that the evaluation of the dependence $\mathrm{EB}(\Delta)$ often
requires much better statistics than that needed for the MSDs presented in
Fig.~\ref{fig-msd-tamsd-out}. The reason is the large spread of $\overline{
\delta^2(\Delta)}$ between trajectories at all lag times $\Delta$, see the
red curves in Fig.~\ref{fig-msd-tamsd-out}. This scatter has more severe
implications on the ergodicity breaking parameter containing the square
$\left<\left(\overline{\delta^2}\right)^2\right>$, see Eq.~(\ref{EB-2D}), and
involving the averaging over $N$ traces.

\begin{figure}
\includegraphics[width=7cm]{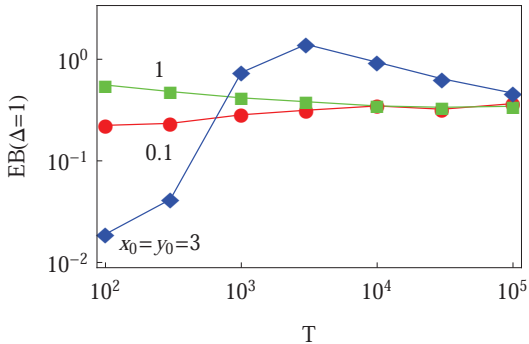}
\caption{Dependence of EB($\Delta=1)$ on the trace length $T$. At least $N=$300
trajectories were used to compute each point in the graph. Parameters are the
same as in Fig.~\ref{fig-msd-tamsd-out}.} 
\label{fig-eb-trace-length}
\end{figure}

\begin{figure}
\includegraphics[width=7cm]{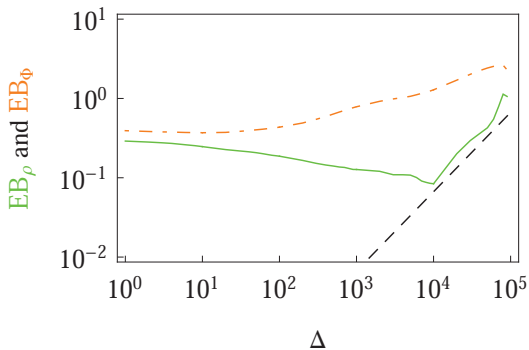}
\caption{Direction-dependent ergodicity breaking parameters. The Brownian asymptote
(\ref{eq-eb-d2}) is represented by the dashed line. Parameter are the same as in
Fig.~\ref{fig-msd-tamsd-out}. The colour coding corresponds to
Fig.~\ref{fig-radius-phase}.} 
\label{fig-eb-dir}
\end{figure}     
     
The dependence of the initial value $\mathrm{EB}(\Delta=1)$ on the trajectory
length $T$ for different initial conditions is illustrated in
Fig.~\ref{fig-eb-trace-length}. Similar to the 1D situation treated in
Refs.~\cite{hdp13a,hdp13b}, the variation of $\text{EB}(T)$ depends on how far
the system is away from the ergodic state for the imposed initial conditions.
For instance, the ergodicity breaking parameter EB$(\Delta=1)$ for $x_0=y_0=3$
and short traces with $T=10^2$ is quite close to the Brownian value given by
Eq.~(\ref{eq-eb-d2}). Conversely, for the same initial conditions but longer
trajectories, $T=10^{3\dots5}$, the system is more non-ergodic and the
corresponding $\mathrm{EB}$ parameter is larger than that for $x_0=y_0=0.1$
or 1 (Fig.~\ref{fig-eb-trace-length}). This is the reason why we observe
intersection of curves for different $\{x_0,y_0\}$ values shown in
Fig.~\ref{fig-eb-trace-length}. At $T\to\infty$ the ergodicity breaking parameter
tends to a universal value.

The data show that system heterogeneities indeed cause a weak ergodicity breaking
in the 2D HDP with diffusivity (\ref{eq-Dr}). Due to the non-equivalence of the
radial and azimuthal diffusion, we predict a direction-dependent ergodicity
breaking parameter, see Fig.~\ref{fig-eb-dir}. We observe that the radial
azimuthal ergodicity breaking parameters $\mathrm{EB}_\rho$ and $\mathrm{EB}_\Phi$
become quite close in the limit $\Delta/T\ll1$. In the limit of long lag times,
as $\Delta\sim T$, the parameter $\mathrm{EB}_\rho$ behaves similarly to
$\mathrm{EB}$ computed from $\mathbf{r}(t)$, compare Figs.~\ref{fig-eb} and
\ref{fig-eb-dir}. In contrast, the azimuthal parameter $\mathrm{EB}_\Phi$ does
not approach the Brownian asymptote (\ref{eq-eb-d2}) at later stages of the time
averaged trajectories.

\subsection{PDF and spreading of particles}

\begin{figure}
\includegraphics[width=8.8cm]{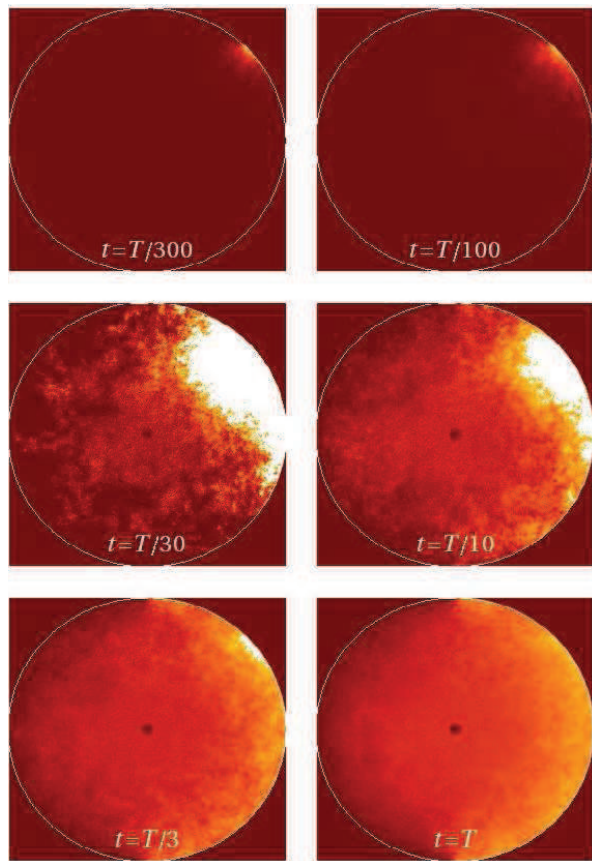}
\caption{Series of PDFs in our 2D `cell' for the temporal spreading of random
walkers starting at the cell boundary at $x_0=y_0=R/\sqrt{2}$. $N=150$
trajectories of $T =10^5$ time steps were analysed. The cell radius is $R=5$, and
the times $t$ of the snapshots are indicated in the panels. The dark spot in the
centre of each graph is due to the faster diffusion at $r=0$ and a finite grid
for sampling and projecting $r(t)$ traces.}
\label{fig-pdf}
\end{figure}

The spreading of particles starting at the cell boundary at $r=R$ is characterised
by the PDF shown in Fig.~\ref{fig-pdf}. The initial accumulation of particles near
the reflecting outer wall in the region of low diffusivity contributes to the
enhanced azimuthal spreading. This spreading remains profound also at later times,
because of the Brownian diffusive behaviour in azimuthal direction, as contrasted
to subdiffusive spreading in the radial direction, see Fig.~\ref{fig-radius-phase}.
The overall trend is similar to the 1D case \cite{hdp13a,hdp13b}, where at long
times the particles tend to accumulate in the regions of lower diffusivity.
Naturally, the average effective jump length of particles diffusing near $r=0$ is
larger than in the region of slow diffusion near the cell boundary. As one can see
from Fig.~\ref{fig-pdf}, a strong azimuthal spread at $t=T/30\dots T/10$ turns
into a profound invasion of particles over the entire cell at $t=T$ (for trace
length $T=10^5$ and $N=150$ analsed trajectories in this figure).

\begin{figure}
\includegraphics[width=7cm]{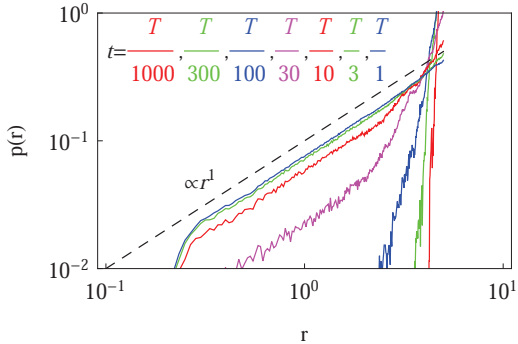}
\caption{Dependence of the radial distribution function $p(r)$ for particle
invasion into a circular nucleus-free domain $0<r<R$ starting from the cell
boundary. Parameters are the same as in Fig.~\ref{fig-pdf}.} 
\label{fig-pr}
\end{figure}

\begin{figure}
\includegraphics[width=8.8cm]{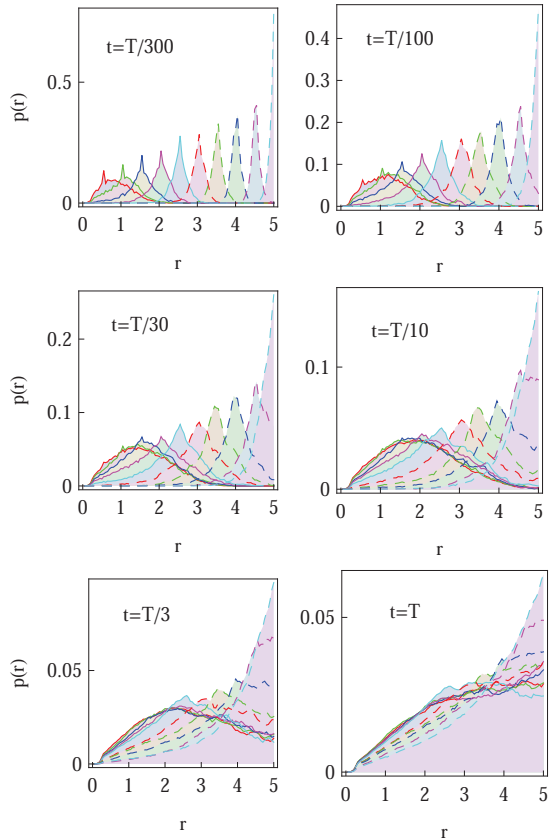}
\caption{Spreading of particles starting initially `uniformly-distributed' at
10 positions within
a circular domain $0<r<R$. The particles appear to focus towards the region of
low diffusivity at $r=R=5$ at longer diffusion times. For each choice of initial
positions $\{x_0,y_0\}$ we generated $N=200$ trajectories of length $T=10^4$.}
\label{fig-pdf-homo}
\end{figure}

The time evolution of the radial PDF shown in Fig.~\ref{fig-pr} quantifies the
2D plots in Fig.~\ref{fig-pdf} when particles are initially released at the
fringe of the cell. We observe that for longer trajectories the maximum of
the PDF, initially localised at $r=R$, progressively spreads and approaches
a universal scaling law given by $p(r)\simeq r$.

We also simulated the diffusive `focusing' of walkers, that were initially
homogeneously distributed in the cell. We find that fast-diffusing particles
leave the region near the origin at $r=0$ relatively quickly and progressively
shift the maximum of the PDF towards the region of slow diffusion near
the cell periphery, see the graphs for different times $t$ in
Fig.~\ref{fig-pdf-homo}. This trend is similar to the 1D situation with
power-law diffusivity \cite{hdp13a}.

\subsection{Survival probability: diffusion from the nucleus to the membrane}

\begin{figure}
\includegraphics[width=7cm]{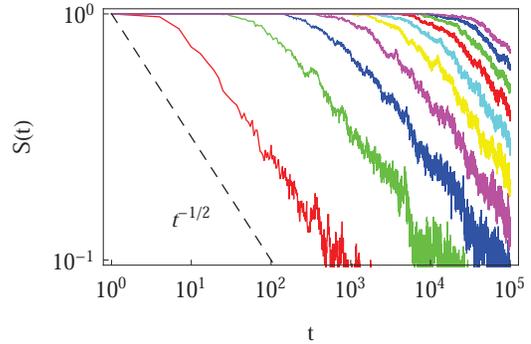}
\caption{Survival probability in the domain of radius $R$ for initial particle
release in the centre of the cell. The universal scaling (\ref{eq-surv-t12}) is
shown by the dashed line. Parameters: $R=1,2,\dots,10$ for the curves from left
to right, $x_0=y_0=0.1$, $T=10^5$, and $N=300$.} 
\label{fig-surv-prob-out}
\end{figure}

After starting at the cell centre and diffusion towards the cell membrane at
$r=R$, the probability of staying in the domain of radius $R$ is described by
the survival probability $S(t)$ shown in Fig.~\ref{fig-surv-prob-out}. The
simulations results obey the universal scaling
\begin{equation}
S(t)\simeq t^{-1/2}.
\label{eq-surv-t12}
\end{equation}
Naturally, for larger cells the diffusing particles start to follow this asymptote
at later times, as it takes longer to reach the outer cell border by diffusion.
Relatively strong variations of $S(t)$ at later times are due to back-and-force
diffusion of individual particles through the outer boundary (which was treated
permeable in the algorithm for computing $S(t)$ in Fig.~\ref{fig-surv-prob-out}).
We have checked that the scaling law (\ref{eq-surv-t12}) is valid also for other
initial positions $\{x_0,y_0\}$ in the cell (results not shown). We also expect
Eq.~(\ref{eq-surv-t12}) to remain valid for other choices of the diffusivity
variation $D(r)$ in the cell.

\begin{figure}
\includegraphics[width=7cm]{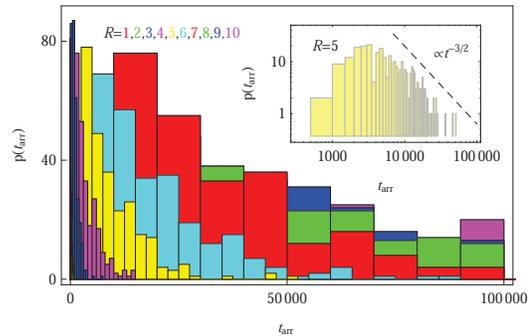}
\caption{Distributions of arrival times to the cell boundary for diffusion of
particles initially released in the cell centre, plotted for varying cell radius
$R$. In the inset we demonstrate the scaling $t^{-3/2}$ expected from the survival
probability in Fig.~\ref{fig-surv-prob-out}. The colour scheme and parameters are
the same as in Fig.~\ref{fig-surv-prob-out}.} 
\label{fig-arrival-out}
\end{figure}

We also computed the distribution of arrival times of particles diffusing from
the cell centre to the cell boundary, see Fig.~\ref{fig-arrival-out}. These
distributions $p(t_{\mathrm{arr}})$ reveal a wide spread, particularly at large
$R$ values, indicating large trajectory-to-trajectory fluctuations. From these
distributions we determine the threshold time $t_{1/2}$ at which 50\% of the
fastest particles reach the outer cell boundary. Such a threshold characteristic
is often important for biological problems, e.g., in the dynamics of population
spreading or proliferation of viral infections.

\begin{figure}
\includegraphics[width=7cm]{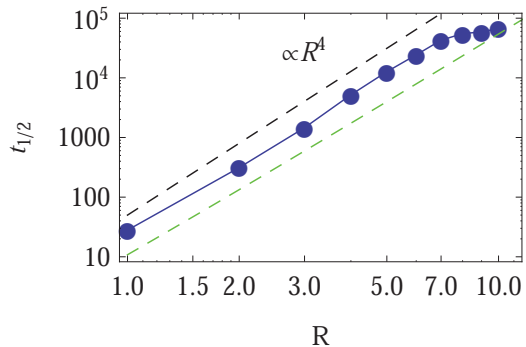}
\caption{Scaling of $t_{1/2}$ for diffusion from centre to cell boundary with cell
size $R$. The asymptotes correspond to Eqs.~(\ref{eq-t12-r4}) (top) and
(\ref{eq-t12-ave}) (bottom line). Parameters are the same as in
Fig.~\ref{fig-arrival-out}.} \label{fig-t12}
\end{figure}

The function $t_{1/2}(R)$ obtained via the analysis of the histograms presented
in Fig.~\ref{fig-arrival-out} often turns out to be bounded by two asymptotes.
The first one is defined via the slowest diffusivity at the cell boundary $r=R$.
Namely, from elementary scaling arguments we can write
\begin{equation}
t_{1/2}(R)\simeq\frac{R^2}{2D(R)}\simeq\frac{R^4}{2A}.
\label{eq-t12-r4}
\end{equation}
The second characteristic time scale is defined via the average diffusion
coefficient in the domain,
\begin{equation}
\label{eq-Dave}
\left<D\right>=\frac{\int_a^R{D(r)rdr}}{(R^2-a^2)/2},
\end{equation}
namely
\begin{equation}
\label{eq-t12-ave}
t_{1/2}(R)\simeq\frac{R^2}{2\left<D\right>}\simeq\frac{R^4}{2A\log[1+R^2/A]}.
\end {equation}
These asymptotes (respectively, the black and green lines in Fig.~\ref{fig-t12})
indicate the leading-order scaling $t_{1/2}\sim R^4$ (apart from the logarithmic
correction).

The time $t_{1/2}$ characterises the arrival of the fastest half of a population
of diffusing walkers, and it can be related to the effectiveness and reliability
of the target search in such a heterogeneous medium. It is particularly important
as the arrival time distributions are skewed \cite{rm-fptd}, compare
Fig.~\ref{fig-arrival-out}. Consequently, the mean of the
distribution and its width are not the best indicators of the arrival statistics
\cite{rm-fptd}, and instead $t_{1/2}$ should be used. In our 2D bounded domain
the first-passage time dynamics and the histograms for $p(t_{arr})$ can be
fitted, e.g., by a generalised Gaussian distribution \cite{rm-fptd,greben12} (not
shown). Clearly, for a larger domain size $R$ the width of the
distributions of arrival times grows, because of accumulated statistical
fluctuations among diffusing particles with longer trajectories, see
Figs.~\ref{fig-arrival-out} and \ref{fig-arrival-in}.

\begin{figure}
\includegraphics[width=7cm]{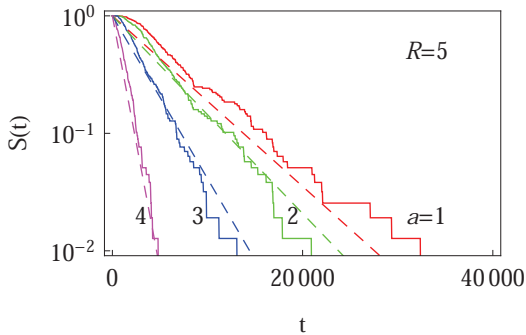}
\caption{Exponential decay of the survival probability for the diffusion of
particles starting at the cell membrane. The dashed lines represent
Eq.~(\ref{eq-exp-tD}). The radii $a$ of the inner absorbing boundary are
indicated in the graph, other parameters are the same as in
Fig.~\ref{fig-surv-prob-out}. } 
\label{fig-surv-prob-in}
\end{figure}

\begin{figure}
\includegraphics[width=7cm]{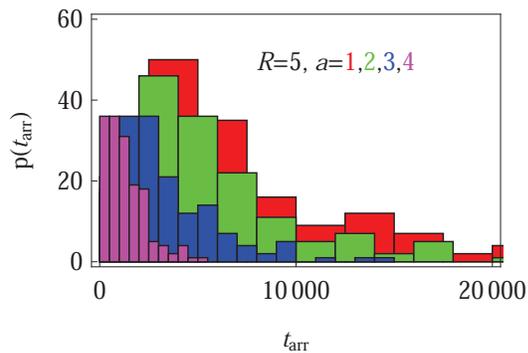}
\caption{Distributions of arrival times from the cell boundary to a nucleus of
radius $a$. Colour coding and parameters are the same as in
Fig.~\ref{fig-surv-prob-in}.} 
\label{fig-arrival-in}
\end{figure}

\subsection{Survival probability: diffusion from the cell membrane to the nucleus}
   
The survival probability for the diffusion from the outer boundary to the cell
`nucleus' exhibits the exponential scaling
\begin{equation}
S(t)\simeq\exp\left(-\frac{t}{t^{\star}}\right),
\label{eq-exp-tD}
\end{equation}
in contrast to the $t^{-1/2}$ law (\ref{eq-surv-t12}) for the opposite direction
in the same domain. This exponential scaling is akin to the standard problem
of 2D diffusion in a circular domain with a sink \cite{redner}, see also
Ref.~\cite{rm-fptd} for the exponential scaling of $S(t)$ in the 3D case.

The characteristic time $t^{\star}$ of the decrease of $S(t)$ with time corresponds
to the time the particles spend diffusing from the outer to the inner boundary in
a medium with average diffusivity. As the radial diffusion is quasi-1D we can
write
\begin{equation}
t^{\star}\sim\frac{(R-a)^2}{2\left<D\right>}.
\end{equation}
For not too large $(R-a)$ values, when the medium diffusivity varies only
moderately in the concentric shell, such an ansatz for $t^{\star}$ works quite
well. These asymptotes are shown as dashed lines in Fig.~\ref{fig-surv-prob-in},
in comparison to the simulation results for $S(t)$.

Similarly to the results for nucleus-to-membrane diffusion in
Fig.~\ref{fig-arrival-out}, we evaluate the distribution of the arrival times
from the cell periphery to the nucleus of different sizes, compare
Fig.~\ref{fig-arrival-in}.

\section{Discussion and Outlook}
\label{sec-discussion}

We studied the diffusion of particles in a 2D circular domain with a radially
varying diffusivity $D(r)$. We showed that the resulting HDP is weakly non-ergodic
in the sense that time and ensemble averages of physical quantities such as the
MSD behave differently. This effect was shown to depend on the initial conditions
of the diffusive walkers. The diffusion in the direction of the diffusivity
gradient was shown to be anomalous, while the azimuthal diffusion occurs in a
nearly constant environment and is Brownian. This behaviour is reminiscent of the
radial and azimuthal diffusion of viral particles monitored in the bacterial
cytoplasm, with purely radially varying diffusivity \cite{brauch01}.

Specifically for the evaluation of single particle tracking data, our results for
the non-ergodicity imply that (i) the time averages of physical quantities such as
the MSD behave differently from their ensemble analogues, and that (ii) individual
time averages are not reproducible, i.e., there occurs a major scatter in the
amplitudes of these quantities. Both need to be taken into account for a proper
physical interpretation of data.

We demonstrated that the diffusion from the domain centre to its boundary (nucleus
to membrane) and the reverse process obey entirely different behaviours for the
respective survival probabilities. Namely, the $S(t)\simeq t^{-1/2}$ scaling law
was found for nucleus-membrane diffusion and the exponential $S(t)\simeq e^{-t/t
^{\star}}$ decay was identified for membrane-nucleus diffusion. This latter fact
as well as the spreading of particles according to these two scenarios can be
rationalised in terms of a domain-averaged diffusion coefficient.

A quantitative understanding
and the ability to tune viral diffusion in living cells has enormous potential
as a tool to control and hopefully suppress the proliferation of infection.
Viral gene delivery carriers \cite{AAV-traffik,buning-therapy} with a high
transfection efficiency actively transported by motors
\cite{gene-carriers-active-transport}
are nowadays extensively used for gene delivery purposes. We note that our
model may also be applied to macroscopic systems. Thus, the spatial spreading
of epidemics in a population of animals subject to non-homogeneous habital
or foraging conditions is another possible area for application for our model.

In the present paper, we focused on the statistical and nonergodic properties
of HDPs in circular domains. A mathematical investigation of the process of viral
infection in the presence of three inter-connected diffusion pathways (anomalous
diffusion, normal diffusion, and active directional transport) is currently
under way \cite{ccr3channels}.

\section{Acknowledgements}

We thank E. Barkai, H. B{\"u}ning, A. Godec, and T. K{\"u}hn for discussions. We
also acknowledge funding from the Academy of Finland (FiDiPro scheme to RM)
and the Deutsche Forschungsgemeinschaft (Grant CH 707/5-1 to AGC).

\end{document}